\documentclass[prl,aps,amssymb,showpacs,twocolumn,superscriptaddress,times,10pt]{revtex4-1}

\usepackage[english]{babel}
\usepackage[pdftex,colorlinks=false]{hyperref}
\usepackage{amsmath,amssymb,amsthm}

\usepackage{amsmath,amssymb,mathrsfs}
\usepackage{latexsym}
\usepackage{graphicx}
\usepackage{epstopdf}
\usepackage{color}
\usepackage{amsfonts}
\usepackage{bm}
\usepackage{bbm}
\usepackage{multirow}
\usepackage{subfiles}

\usepackage{natbib}

\usepackage{ulem} 

\usepackage{tikz} 
\usepackage{tikz-feynman}

\usepackage{mathtools} 
\usepackage{physics}
\usepackage[version=3]{mhchem}

\usepackage{microtype} 
\usepackage{csquotes} 
\usepackage{booktabs} 

\makeatletter
\renewcommand\@make@capt@title[2]{%
	\@ifx@empty\float@link{\@firstofone}{\expandafter\href\expandafter{\float@link}}%
	{\textbf{#1}}\@caption@fignum@sep#2\quad
}%
\makeatother

\makeatletter
\g@addto@macro\bfseries{\boldmath} 
\makeatother

\usepackage{graphicx}
\usepackage{placeins}

\usepackage{siunitx} 
\DeclareSIUnit\year{a}
\DeclareSIUnit\pixel{px}
\DeclareSIUnit\line{line}

\usepackage{scrlayer-scrpage} 
\usepackage{lastpage} 
\usepackage{placeins}

\newcommand{\cre}{{\dag}}
\newcommand{\ann}{{\vphantom{\dag}}}
\newcommand{\Tc}{T_{{\rm c}}}

\newcommand{\mperiod}{\,\text{.}} 
\newcommand{\mcomma}{\,\text{,}}

\graphicspath{./figures/}

\begin{document}

\title{Chiral surface superconductivity in half-Heusler semimetals}

\author{Tilman Schwemmer}\email{tilman.schwemmer@physik.uni-wuerzburg.de}
\affiliation{Institute for Theoretical Physics, University of Wuerzburg, D-97074 Wuerzburg, Germany}

\author{Domenico Di Sante}\email{domenico.disante@unibo.it}
\affiliation{Department of Physics and Astronomy, University of Bologna, 40127 Bologna, Italy}
\affiliation{Center for Computational Quantum Physics, Flatiron Institute, 162 5th Avenue, New York, NY 10010, USA}

\author{J\"org Schmalian}\email{joerg.schmalian@kit.edu}
\affiliation{Institute for Theory of Condensed Matter and Institute for Quantum Materials and Technologies,
Karlsruhe Institute of Technology, Karlsruhe 76131, Germany}

\author{Ronny Thomale}\email{rthomale@physik.uni-wuerzburg.de}
\affiliation{Institute for Theoretical Physics, University of Wuerzburg, D-97074 Wuerzburg, Germany}

\begin{abstract}
We propose the metallic and weakly dispersive surface states of half-Heusler semimetals as a possible
domain for the onset of unconventional surface superconductivity ahead
of the bulk
transition.
Using density functional theory (DFT) calculations and the random
phase approximation (RPA), we analyse the surface band structure of
\ce{LuPtBi} and its propensity towards Cooper pair formation induced
by screened electron-electron interactions in the presence of strong
spin-orbit coupling.
Over a wide range of model parameters, we find an energetically favoured chiral superconducting condensate featuring Majorana edge modes, while low-dimensional order
parameter fluctuations trigger time-reversal symmetry breaking to precede the superconducting transition.
\end{abstract}

\date{\today}
\maketitle


\textit{Introduction.}---
The analysis of many established unconventional superconductors has been predominantly
focused on the bulk limit of quasi-two-dimensional (quasi-2D) layered materials such as the
cuprates, ruthenates, or certain families of iron-based
superconductors~\cite{Proust2019,Si2016,Stewart2011,Wen2011,Mazin2009,Mackenzie2017}.
This fact may be understood in terms of the generically enhanced nesting features
of quasi-2D Fermi surfaces compared to three-dimensional isotropic
bulk systems, which plays into the spin fluctuation paradigm of
unconventional pairing. It does, however, provoke the question whether
particularly engineered quasi-2D surface bands of 3D bulk crystals might provide an
alternative arena for unconventional superconductivity (SC).

While interface SC between insulating oxide
materials such as \ce{LaAlO3}/\ce{SrTiO3} or \ce{KTaO3} is
well established by now~\cite{Reyren2007,Bert2011,Gozar2008,Li2011,Liu2021},
any sensible distinction between surface and bulk SC
derived from a single 3D bulk (semi)metal appears to be rare.
This is because the SC 3D bulk transition is expected to set the decisive scale due to its generically 
dominant density of states.
As we consider unconventional pairing, however, such a view requires revision.
There, it is conceivable that the interaction over bandwidth
ratio and the nesting of surface bands can be a superior driver of
unconventional pairing when compared to the bulk bands,
and that a SC surface transition could precede the bulk transition.
Evidence of surface SC has been reported in a variety of 3D semimetals
where both topological and non-topological (such as of Shockley or Tamm
type) surface states form rather flat bands~\cite{Shvetsov2019, Xing2020, Shen2020, Song2021, Liu2022}.

In this Letter, we investigate the possible nature of unconventional
surface SC in half-Heusler semimetals, and particularize our analysis
to LuPtBi.
\ce{LnPtBi} (\ce{Ln}=\ce{Y},\ce{Lu}), with a reported bulk \(\Tc\) of
\(\approx \SI{1}{\kelvin}\), has been highlighted as a candidate for
unconventional SC.
This is because aside from strong spin-orbit coupling (SOC), a non-symmorphic
crystal structure, and the existence of a quadratic band touching point near
the Fermi energy~\cite{Butch2011, Bay2012, Tafti2013, Majumder2019}, the
electron-phonon coupling appears too low to explain the observed
$\Tc$~\cite{Meinert2016}.
Several avenues of exotic unconventional 3D bulk SC have been proposed
for \ce{LuPtBi} and \ce{YPtBi} \cite{Brydon2016, Timm2017, Savary2017,
Boettcher2018, Wang2018, Bahari2022}.
One central motif in these proposals is the SOC, where the associated enlarged
crystal double group irreducible representations may allow the formation of
SC out of higher-spin fermionic quasiparticles some confirmation of
which is claimed to be observed for \ce{LuPdBi}~\cite{Ishihara2021}.
In the following, we will take a different path to a possible explanation of
the nature of SC in half-Heuslers.
Note, that the (111) surfaces of these half-Heuslers feature rather flat
metallic surface states, and hence might also be prone to surface SC.
Combined with reports indicating that a small fraction of its electronic
density of states undergoes an SC transition
at \(\Tc \approx \SI{6}{\kelvin}\) \cite{Banerjee2015},
we propose that the particularly flat surface bands for
\ce{LuPtBi}'s Bi(111) termination could trigger an initial surface SC
transition \(\Tc^{\text{surface}}\),
preceding the bulk SC transition at some lower \(\Tc^{\text{bulk}}\).
Following this assumption, we find that the preferred unconventional SC
pairing function transforms under the \(E\) irreducible
representation of the surface's C\(_{3v}\) point group in the presence
of strong Kane-Mele type and Rashba type SOC.
Our mean field analysis then suggests that the SC surface condensate
spontaneously breaks time-reversal symmetry (TRS).

\begin{figure}[!t]
	\includegraphics[width=1\columnwidth]{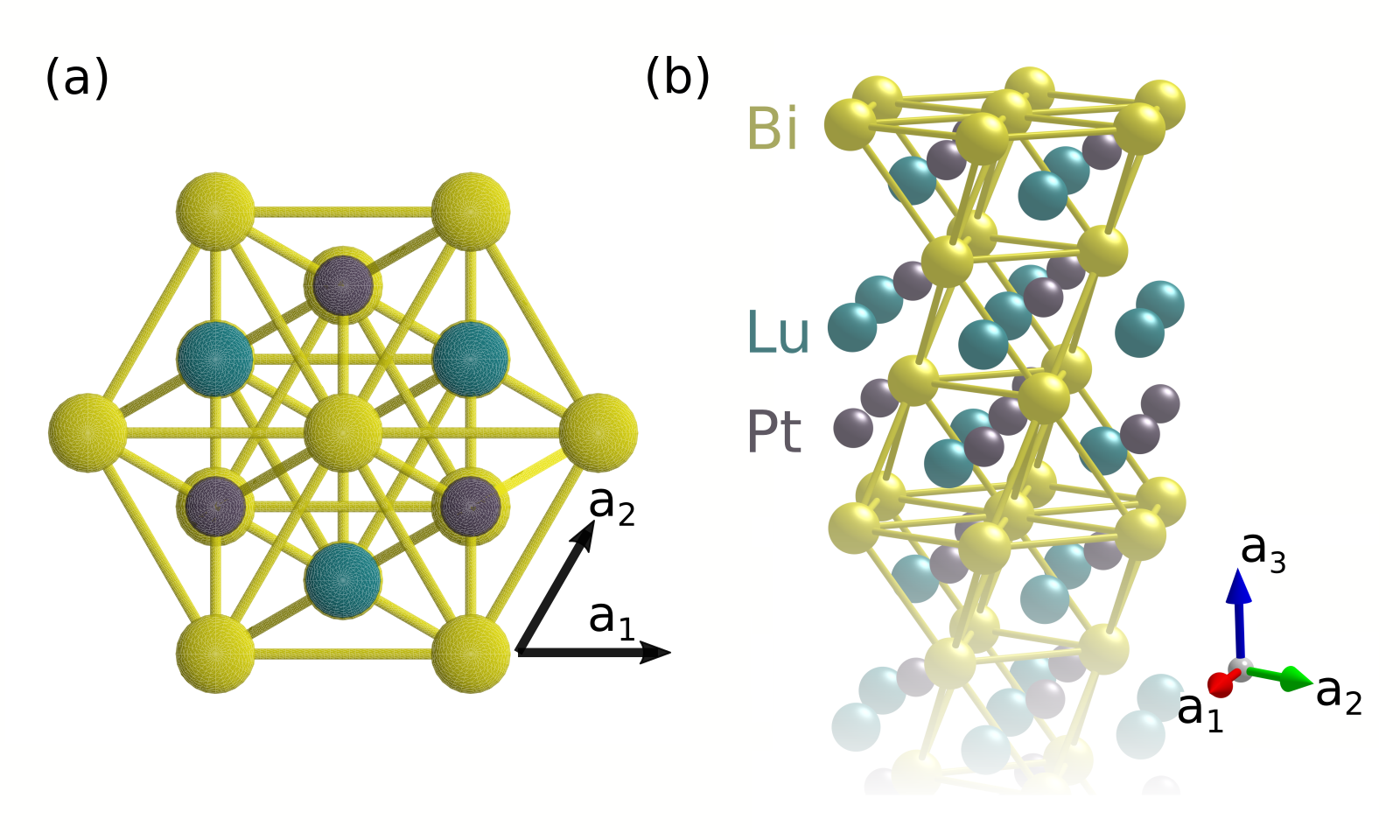}
	\caption{\label{fig:crystal_structure} (a) Surface crystal
          structure formed by a triangular lattice of Bi atoms
          whose C\(_{6v}\) point group symmetry is broken down to
          C\(_{3v}\) by the presence of Lu and Pt atoms in the layers
          below. (b) Bulk crystal structure of \ce{LuPtBi} with the
          non-symmorphic nature of the half-Heusler crystal structure
          along the (111) direction ($c$-axis).}
\end{figure}


\textit{Ab-initio modeling.}---
LuPtBi is a half-Heusler compound and crystallizes in the non-symmorphic space
group (SG) F-43m (\# 216)~\cite{Canfield1991, Tafti2013}.
Since we are interested in the surface states along the 111 direction,
a non-primitive setting for the unit cell with broken translational symmetry
in the \(\pmb{e}_z=\pmb{a}_3\) direction was chosen, yielding
the two dimensional space group P3m1 (SG \# 156 = plane group \# 14).
The \(\Gamma\)-point of this SG is invariant under the C\(_{3v}\) point group,
despite the fact that the Bi atoms at the surface form a triangular lattice.
This becomes clear upon considering \autoref{fig:crystal_structure}\,(a)
and realizing that, while the system is invariant under a mirror operation along
the \(\pmb{a}_1+\pmb{a}_2\) direction, it exchanges the positions of Lu and Pt
under the \(\pmb{a}_2-\pmb{a}_1\) mirror, resulting in a breaking of the
C\(_{6v}\) symmetry of the triangular Bi lattice.

\begin{figure}[t]
	\includegraphics[width=1\columnwidth]{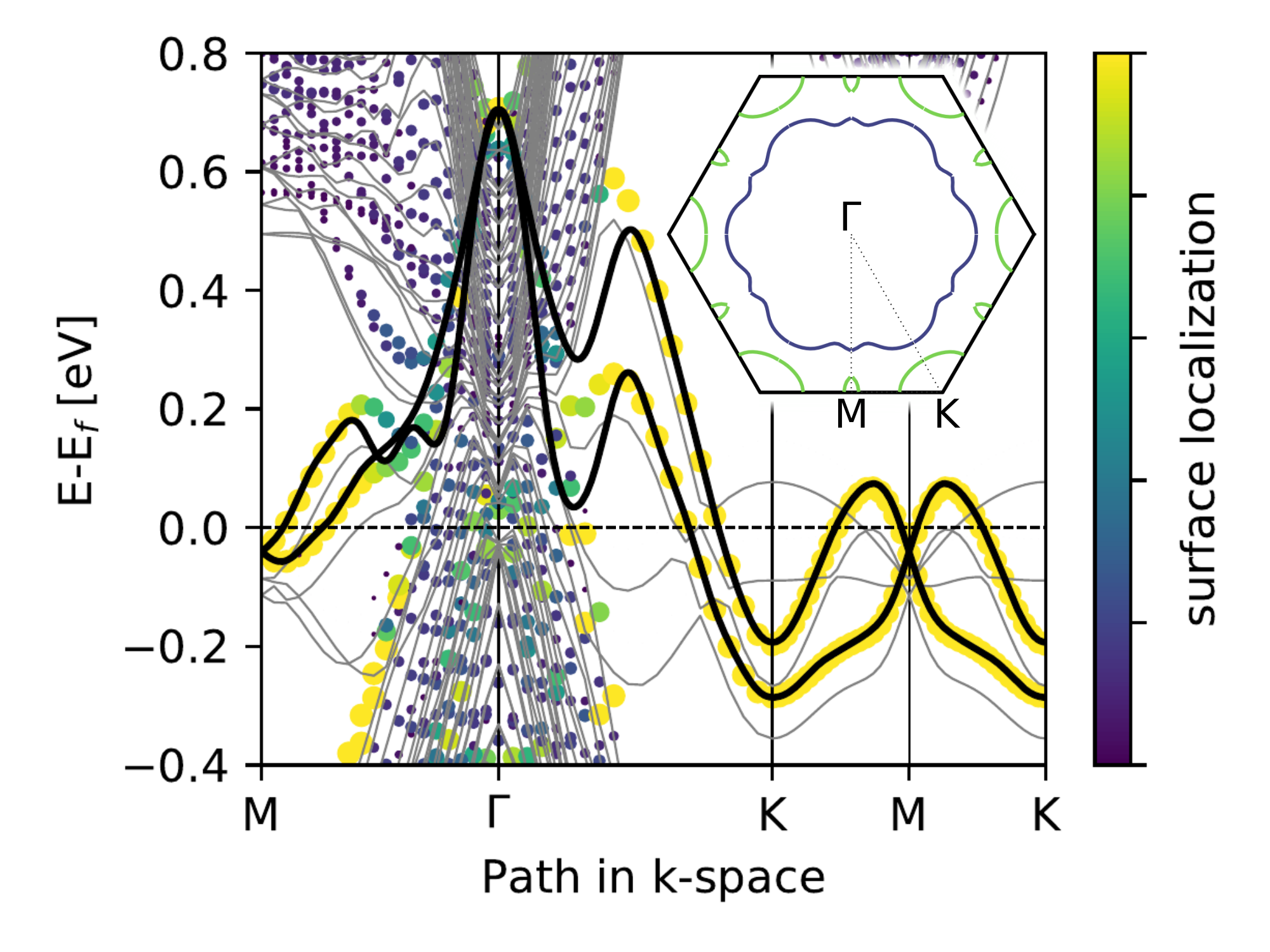}
	\caption{\label{fig:band_structure}
	The resulting surface band structure as obtained by DFT as thin grey lines
	with the overlaid dots indicating the surface localization in the calculation.
	The continuous black lines represent a fit to the calculation obtained by
	using maximally localized Wannier functions on the Bi surface.
  Bands without any overlaid points are localized on the opposite surface
	with a different (non-Bi) termination.
	The inset shows the resulting Fermi surface and highlights the two
	helicity sheets in different colours.}
\end{figure}

From the DFT calculation of a slab of material shown in~\autoref{fig:band_structure}
we find that the least dispersive low-energy bands emerge from
dangling bonds of the Bi p\(_z\) orbitals at the crystal's
surface.
Furthermore, we find that they are likewise most suited for unconventional
pairing in terms of Fermiology and density of states at the Fermi level,
which is why we restrict our analysis to these bands.

We use the framework of maximally localized Wannier functions in order to
obtain a best fit tight binding model, using a \textit{single}
Bi p\(_\text{z}\) orbital on the 1a Wyckoff position of the P3m1 space group.
\autoref{fig:band_structure} shows the resulting band structure
overlaid as a black line over the \textit{ab-initio} data.
The resulting tight-binding Hamiltonian yields
\begin{equation}\label{eq:H0}
	\hat{H}_0 = \sum_{\pmb{k}} \psi_{\pmb{k}}^\dag (\epsilon(\pmb{k}) + \pmb{g}(\pmb{k})\pmb{\sigma}) \psi_{\pmb{k}}^{\phantom{\dag}}
	= \sum_{\pmb{k}} \psi_{\pmb{k}}^\dag h(\pmb{k}) \psi_{\pmb{k}}^{\phantom{\dag}}\mcomma
\end{equation}
with the combined creation operator
\begin{equation}
	\psi_{\pmb{k}}^\dag = (c_{\pmb{k},\uparrow}^\dag\text{ , }c_{\pmb{k},\downarrow}^\dag)\mcomma
\end{equation}
for up (down) spin electrons \(c_{\pmb{k}\uparrow}^\dag\)
(\(c_{\pmb{k}\downarrow}^\dag\)) at momenta \(\pmb{k}\). All details
on \(\epsilon(\pmb{k})\) and \(\pmb{g}(\pmb{k})\) are provided in~\footnote{Supplemental Material and references \cite{Kresse1996, Bloechl1994, Perdew1996, Mostofi2008, Franchini2012} therein. Additionally, relevant input and output data of the first-principles study can be downloaded from \url{https://doi.org/10.5281/zenodo.7352135}}.

The strong SOC of Bi, modelled by \(\pmb{g}(\pmb{k})\) in \autoref{eq:H0},
results in a significant splitting of the bands due to the broken inversion
symmetry on the crystal's surface.
Kramer's theorem protects a two-fold degeneracy of the band at the TR
invariant momenta \(\Gamma\) and \(M\), with a linear dispersion in the vicinity
of the latter.
We label the system's energy eigenstates with their helicity eigenvalue
\(\lambda=\pm 1\), where the helicity operator \(\Lambda_{\pmb{k}}\)
is defined as the normalized SOC term in the Hamiltonian
\begin{equation}
	\Lambda_{\pmb{k}} c_{\pmb{k}\lambda}^\dag \ket{0} =
	\frac{\pmb{g}(\pmb{k})\pmb{\sigma}}{|\pmb{g}(\pmb{k})|} c_{\pmb{k}\lambda}^\dag \ket{0} =
	\lambda c_{\pmb{k}\lambda}^\dag \ket{0} \mperiod
\end{equation}
Using this operator, we define a unitary transformation
\(u_{\lambda,\sigma}(\pmb{k})\) diagonalizing \autoref{eq:H0}
\begin{equation}
	c_{\pmb{k},\lambda}^\cre = \sum_{\sigma} u_{\lambda,\sigma}(\pmb{k}) c_{\pmb{k},\sigma}^\cre
\end{equation}
at each momentum.
The colour of the Fermi surface pockets depicted in the inset of \autoref{fig:band_structure}
labels the helicity eigenvalue of different points on the Fermi surface, also
indicating the opposite chirality of the spin texture of the corresponding pockets.
Both our band structure and Fermi surface are in good agreement with the ARPES
measurements presented in~\cite{Liu2016}~\footnote{The small Fermi surface
pocket centred around \(\Gamma\) seen in experiment is dominated by strongly
dispersing Bi p\(x\) and p\(_y\) orbitals.
While it is well captured by our DFT calculation, we neglect its contribution
to the surface low-energy model due to its small density of states at the
Fermi level.}.

\begin{figure}
	\includegraphics[width=1\columnwidth]{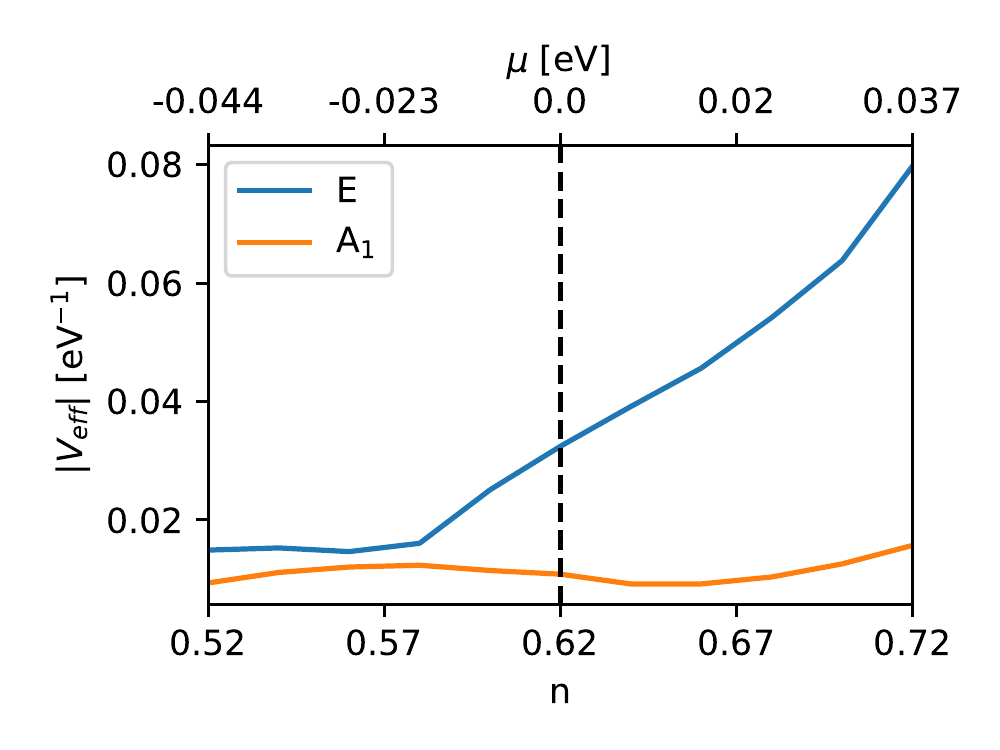}
	\caption{\label{fig:scan_filling} Superconducting pairing strength
	\(V_{\text{eff}}\) in the RPA as a function of the particle occupation
	\(n\) in the surface band.
	While this calculation was done for an intermediate value of
	\(U_0=\SI{0.3}{\eV}\), all qualitative features of this figure are preserved
	upon lowering \(U_0\) to the analytically controlled limit \(U_0 \rightarrow 0\).
	The filling of \(n=0.62\) corresponds to the DFT result, while
  our finding is rather insensitive to the variation of filling.}
\end{figure}

\textit{Unconventional superconductivity.}---
In order to analyse the prevalent superconducting instabilities, we introduce
a suitable basis of Cooper pair states.
Since a parametrization in terms of singlet and triplet states is not
feasible due to the lack of inversion symmetry, we consider pairs of electrons
connected by the TR operator \(\hat{T}=i \sigma_y\mathcal{K}\).
Neglecting inter-band pairing amplitudes, we define
\begin{align}
	\Psi^\dag_{\text{pair}}(\pmb{k};\lambda) &=
	c_{\pmb{k}\lambda}^\dag T c_{\pmb{k}\lambda}^\dag T^{-1} \\
	&=
	\sum_{\sigma,\sigma'}
	\psi_{\lambda,\sigma \sigma'}(\pmb{k})
	c_{\pmb{k}\sigma}^\dag
	c_{-\pmb{k}\sigma'}^\dag
	\mcomma
\end{align}
where
\begin{align}
    \psi_{\lambda,\sigma \sigma'}(\pmb{k}) =
    \sum_{\sigma''}
	u_{\lambda,\sigma}(\pmb{k})
	u^{*}_{\lambda,\sigma''}(\pmb{k})
	(i \sigma_y)_{\sigma''\sigma'}
	\mperiod
\end{align}
Note that our setup allows us to circumvent any issues arising from
the global gauge freedom of \(u_{\lambda,\sigma}(\pmb{k})\) at different momenta
that would arise from choosing \(c_{\pmb{k}\lambda}^\dag c_{-\pmb{k}\lambda}^\dag\)
as the basis for the pairing wavefunction.
Using this parametrization as a starting point, we calculate the system's
propensity towards different superconducting instabilities.
We first implement a perturbative interaction term for our model Hamiltonian
\begin{equation}
	\hat{H} = \hat{H}_0 + U_0 \sum_{\pmb{k}}
	c_{\pmb{k},\uparrow}^\cre
	c_{\pmb{k},\downarrow}^\cre
	c_{\pmb{k},\downarrow}^\ann
	c_{\pmb{k},\uparrow}^\ann \mcomma
\end{equation}
and calculate how it is screened by particle-hole excitations.
In contrast to systems with \(\text{SU}(2)\) symmetry, the bare particle-hole
(PH) susceptibility of our model cannot be reduced to a single momentum
dependent scalar.
Instead we define the generalized PH susceptibility
\begin{equation}
    \chi^0_{\{\sigma_i\}}(\pmb{q},\tau) = \sum_{\pmb{l}} \langle
    \hat{T}_{\tau}
    c_{\pmb{l} + \pmb{q},\sigma_2}^\cre(\tau)
    c_{\pmb{l},\sigma_1}^\ann(\tau)
    c_{\pmb{l},\sigma_3}^\cre(0)
    c_{\pmb{l} + \pmb{q},\sigma_0}^\ann(0)
    \rangle
\end{equation}
with the imaginary time-ordering operator \(\hat{T}_{\tau}\).
We calculate its Fourier transform in the static limit using standard
Matsubara summation techniques for the frequency integral and a discretized
mesh containing \(N_i=1200^2\) points for the
momentum space summation~\footnote{We have checked the convergence of our
results at zero temperature by comparison with integration meshes containing
\(N_i=2000\) points per direction.}.
In the next step, we include the effect of electronic correlations and find
that a bare interaction of \(U_0=\SI{0.37}{\eV}\) is sufficient to drive the
system into an incommensurate magnetic phase with ordering vector
\(\pmb{Q}=[(1+\delta)\pmb{b}_1 + (1-\delta)\pmb{b}_2 ] / 2\) on the RPA level
(\(\delta\approx0.016\)).
Below this critical value, the spin and charge fluctuations induce a
superconducting phase transition~\cite{Scalapino1986,Graser2009,Kemper2010}.
We calculate the Cooper pair scattering amplitudes in our basis
\begin{align}
	\begin{split}
	\Gamma_{\lambda \lambda'}(\pmb{k},\pmb{q}) =
	\sum_{\{\sigma_i\}}
	\psi^{\text{pair}}_{\lambda',\sigma_2 \sigma_3}(\pmb{q})
	\Gamma_{\{\sigma_i\}} (\pmb{k},\pmb{q})
	\psi^{\text{pair}}_{\lambda,\sigma_0 \sigma_1}(\pmb{k}) \mcomma
	\end{split}
\end{align}
within the fluctuation exchange formulation generalized for
spin-orbit coupled systems
\begin{equation}
  \begin{split}
	\Gamma_{\{\sigma_i\}} (\pmb{k},\pmb{q}) &=
	\Gamma^{(1)}_{\{\sigma_i\}} +
	\Gamma^{(2)}_{\{\sigma_i\}} (\pmb{k},\pmb{q}) \mcomma \\
	\Gamma^{(1)}_{\{\sigma_i\}} &=
	U_0 (\delta_{\sigma_0\sigma_2}\delta_{\sigma_1\sigma_3} -
	\delta_{\sigma_0\sigma_3}\delta_{\sigma_1\sigma_2}) \quad\text{and} \\
	\Gamma^{(2)}_{\{\sigma_i\}} &=
    \sum_{\{\tilde{\sigma}_i\}}
    \Gamma^{(1)}_{\sigma_0\tilde{\sigma}_2\sigma_2\tilde{\sigma}_1}
    \chi^{\text{RPA}}_{\{\tilde{\sigma}_i\}} (\pmb{k}-\pmb{q})
    \Gamma^{(1)}_{\tilde{\sigma}_3\sigma_1\tilde{\sigma}_0\sigma_3}
    \\
	&- \sum_{\{\tilde{\sigma}_i\}}
    \Gamma^{(1)}_{\sigma_0\tilde{\sigma}_2\sigma_3\tilde{\sigma}_1}
    \chi^{\text{RPA}}_{\{\tilde{\sigma}_i\}} (\pmb{k}+\pmb{q})
    \Gamma^{(1)}_{\tilde{\sigma}_3\sigma_1\tilde{\sigma}_0\sigma_2}
    \mperiod
  \end{split}
\end{equation}
The generalized RPA susceptibilities
\(\chi^{\text{RPA}}_{\{\sigma_i\}}(\pmb{Q})\) are calculated via the usual
Dyson series resummation of all RPA diagrams, and the momenta
\(\pmb{k}\) and \(\pmb{q}\) are chosen from \(N_p=360\)
unique Fermi surface points~\cite{Duerrnagel2022}.
The resulting Cooper pair scattering amplitude then serves as an input for
the linearised superconducting gap equation
\begin{equation}
    V_{\text{eff}} \Delta_\lambda(\pmb{k}) =  \sum_{\pmb{q}} \Gamma_{\lambda \lambda'}(\pmb{k},\pmb{q}) \Delta_{\lambda'}(\pmb{q}) \mcomma
\end{equation}
whose solutions indicate the leading superconducting gap functions $\Delta$
which can be analysed in terms of irreducible representations of the model's
point group.
The most negative $V_{\text{eff}}$ indicates the maximum paring strength.

\begin{figure*}
	\includegraphics[width=2\columnwidth]{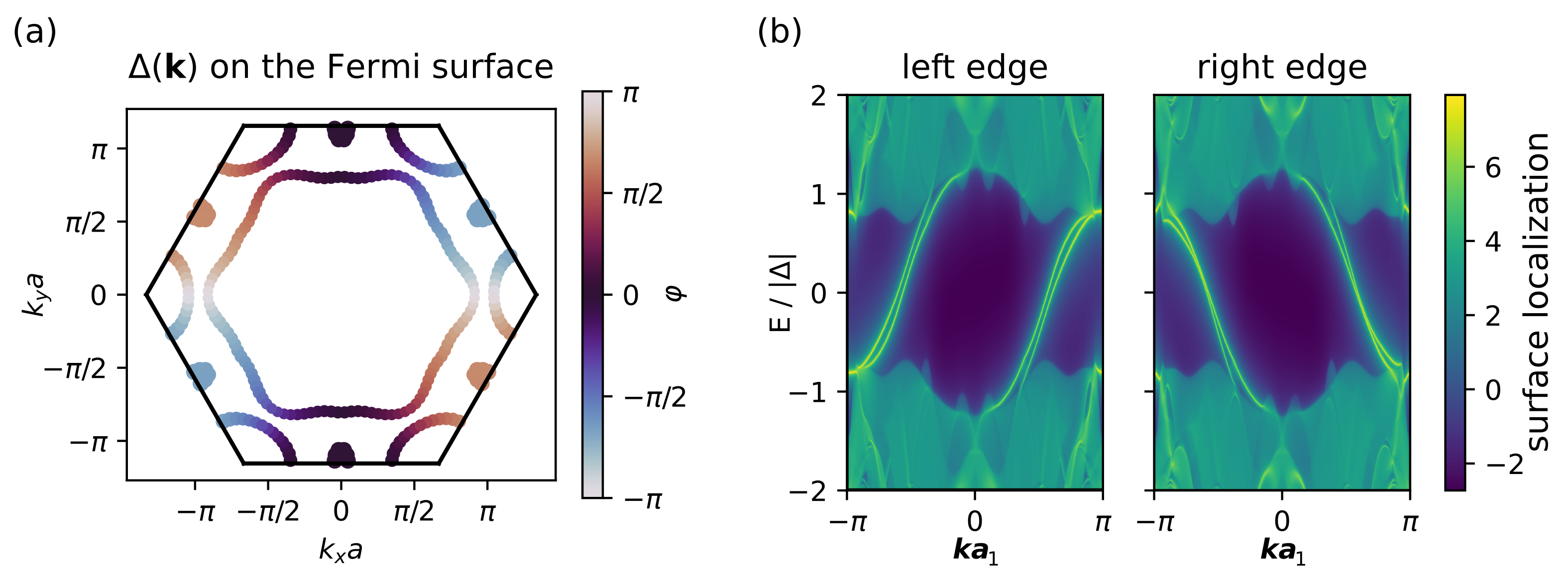}
	\caption{\label{fig:leading_instability}
	(a) The gap function
	\(\Delta(\mathbf{k}) = |\Delta(\mathbf{k})|e^{i\varphi(\mathbf{k})}\)
	of the leading superconducting instability along the Fermi surface.
	The point size is chosen proportional to the absolute value of the gap size
	$|\Delta|$ while the colour represents the phase $\varphi(\mathbf{k})$ of the
	complex condensate.
	One can clearly see the approximately uniform magnitude of the
	chiral $d+id$ pairing as well as the phase winding of $4\pi$.
	(b) Excitation spectrum of the resulting Bogoliubov-de-Gennes Hamiltonian
	on a ribbon with periodic (open) boundary conditions along the
	\(\pmb{a}_2\) (\(\pmb{a}_1\)) directions.
	We show the localization at the left and right boundaries of the ribbon
	respectively in order to demonstrate their different spectra due to the
	C\(_{3v}\) point group of the surface state.}
\end{figure*}

\textit{Results.}---
For our model of Bi (111)-terminated surface states in \ce{LuPtBi}, the leading
eigenvalue of the SC susceptibility is two-fold degenerate over the complete
considered range of doping and relates to an SC gap function transforming
under the \(E\) irreducible representation of
C\(_{3v}\)~\autoref{fig:scan_filling}.
The corresponding gap functions are found to be well approximated by
\begin{equation}
	\begin{split}
		\Phi^{E}_1 &= \frac{2}{\sqrt{3}} \bigg[ \cos(k_x) - \cos(k_x/2) \cos(\sqrt{3}  k_y/2) \bigg] \\
		\Phi^{E}_2 &= \cos(k_x/2+\sqrt{3}  k_y/2) - \cos(k_x/2-\sqrt{3}  k_y/2) \mcomma
	\end{split}
\end{equation}
highlighting the nearest-neighbour character of SC pairing.
The absence of inversion symmetry on the \ce{LuPtBi} surface allows for
mixing of spin-singlet and spin-triplet Cooper pairs, which we can analyse by
comparing the fitted harmonics for the different helicity Fermi surfaces.
From the almost identical fitting parameters for both helicities we infer
the Cooper pair wavefunction to be well approximated by pure spin-singlet pairs.
Given that the filling of a surface state may be sensitive to adsorbates or
other influences that are not present for bulk states, we intentionally
ensure that our findings are stable across a wide range of filling fractions.
In order to discriminate between all possible linear combinations
\(\alpha\Phi^{E}_1 + \beta\Phi^{E}_2\) (with identical pairing potential),
we compute the free-energy \(F\) of the resulting superconducting state.
We find that \(F\) is minimized for the chiral, TRS breaking linear
combinations \(\Phi^{E}_1 \pm i\Phi^{E}_2\)~\autoref{fig:leading_instability},
which yields a full excitation gap across the Fermi surface and therefore
a maximum gain in condensation energy~\cite{Kiesel2012, Nandkishore2012}.
The emergence of a superconducting state with TRS breaking in the surface
can also be understood in terms of a general selection rule for Cooper pair
formation in systems with broken inversion symmetry~\cite{Scheurer2017},
which applies independently of the inclusion of triplet-singlet mixing.
Whenever inversion symmetry breaking of the surface states is strong,
in the sense that the energy scale of the Rashba splitting is larger
than the amplitude of the pairing gap,  multi-component order
parameters  enter in combinations that break TRS.
This conclusion is not only valid near the transition temperature but was shown
to be correct up to arbitrary order of the Ginzburg-Landau expansion.
The alternative of a nematic superconductor that would break a rotation
symmetry in the surface is always energetically more costly~\cite{Scheurer2017}.
The superconducting condensate formed by the \(\Phi^{E}_1 \pm i\Phi^{E}_2\)
combination is characterized by a topological Chern number
$\mathcal{C} = \pm 2$.
One experimentally signature of this topological invariant might hence be
the universal thermal and spin Hall conductivity due to 2 low-energy
topological edge modes~\cite{Read2000,Senthil1999,Horovitz2003},
as highlighted in~\autoref{fig:leading_instability}\,(b).

\textit{Low-dimensional fluctuations.}--- For surface SC, the role of
fluctuations of the SC order parameter are non-negligible.
On the one hand, one expects at $\Tc$ a Kosterlitz-Thouless
transition to a pairing state with finite stiffness.
On the other hand, TRS breaking implies that the relative sign of
$\Phi_{1}^{E}$ and $i\Phi_{2}^{E}$ is tied to an underlying $\mathbb{Z}_{2}$
or Ising degree of freedom.
While SC and TRS breaking occur simultaneously at the mean field level,
fluctuations tend to split the transition~\cite{Hecker2018, Fernandes2019}.
Then, TRS is broken at an Ising phase transition already above $\Tc$~\footnote{More details are provided in the Supplemental Material and references \cite{Polyakov1975, Friedan1980, Orth2014} therein}.
As a result, measurements of the Kerr effect are expected to detect broken TRS even
before a finite stiffness is  detected.
The effect is particularly relevant for systems with low carrier concentration,
as is the case here.
This vestigial TRS breaking is the counterpart of the nematic order above
$\Tc$, observed in doped Bi$_{2}$Se$_{3}$~\cite{Cho2020}.


\textit{Conclusion.}--- We have formulated a theory to reconcile LuPtBi as an instance of an unconventional chiral topological surface
superconductor, which lends itself to several tangible experimental tests.
First, in addition to the possible resolution of the topological edge modes
or vortex spectra, we would expect a finite Kerr signal
not only below $\Tc$, but also above $\Tc$ due to low-dimensional
order parameter fluctuations.
Second, any thermodynamic measurement sensitive to distinguish sub-extensive
surface contributions from bulk contributions should be able to discriminate
between a leading surface transition and a subsequent bulk transition.
While it is likely that our theory will not universally apply to all
half-Heuslers for all surface terminations, it might be promising to reconcile
instances of hitherto observed SC in currently known half-Heuslers from the
viewpoint of unconventional surface superconductivityj, and inspire further efforts into refined sample growth of Half-Heuslers~\cite{https://doi.org/10.48550/arxiv.2211.13106}.


\textit{Acknowledgments.}--- We thank C. Felser and A. Kapitulnik for fruitful discussions. This work is funded by the Deutsche Forschungsgemeinschaft (DFG, German Research Foundation) through Project-ID 258499086 - SFB 1170, through the W\"{u}rzburg-Dresden Cluster of Excellence on Complexity and Topology in Quantum Matter-ct.qmat Project-ID 390858490 - EXC 2147, and through the SFB-TRR 288 Elasto-Q-Mat, Project ID 422213477 (project A07).
The research leading to these results has received funding from the European Union’s Horizon 2020 research and innovation programme under the Marie Sk{\l}odowska-Curie Grant Agreement No. 897276 (BITMAP).
We further gratefully acknowledge the Gauss Centre for Supercomputing e.V. (www.gauss-centre.eu) for providing computing time on the GCS Supercomputer SuperMUC at Leibniz Supercomputing Centre (www.lrz.de).
T.S. acknowledges the hospitality by the Center for Computational Quantum Physics of the Flatiron Institute, where part of the work has been done. The Flatiron Institute is a division of the Simons Foundation.

\bibliographystyle{prsty}
\bibliography{main_and_SM}

\begin{thebibliography}{10}

\bibitem{Proust2019}
C. Proust and L. Taillefer, Annual Review of Condensed Matter Physics {\bf 10},
   409  (2019).

\bibitem{Si2016}
Q. Si, R. Yu, and E. Abrahams, Nature Reviews Materials {\bf 1},  1  (2016).

\bibitem{Stewart2011}
G.~R. Stewart, Reviews of Modern Physics {\bf 83},  1589  (2011).

\bibitem{Wen2011}
H.-H. Wen and S. Li, Annual Review of Condensed Matter Physics {\bf 2},  121
  (2011).

\bibitem{Mazin2009}
I.~I. Mazin and J. Schmalian, Physica C: Superconductivity {\bf 469},  614
  (2009).

\bibitem{Mackenzie2017}
A.~P. Mackenzie, T. Scaffidi, C.~W. Hicks, and Y. Maeno, npj Quantum Materials
  {\bf 2},  40  (2017).

\bibitem{Reyren2007}
N. Reyren {\it et~al.}, Science {\bf 317},  1196  (2007).

\bibitem{Bert2011}
J.~A. Bert {\it et~al.}, Nature Physics {\bf 7},  767  (2011).

\bibitem{Gozar2008}
A. Gozar {\it et~al.}, Nature {\bf 455},  782  (2008).

\bibitem{Li2011}
L. Li, C. Richter, J. Mannhart, and R.~C. Ashoori, Nature Physics {\bf 7},  762
   (2011).

\bibitem{Liu2021}
C. Liu {\it et~al.}, Science {\bf 371},  716  (2021).

\bibitem{Shvetsov2019}
O.~O. Shvetsov, V.~D. Esin, A.~V. Timonina, N.~N. Kolesnikov, and E.~V.
  Deviatov, Physical Review B {\bf 99},  125305  (2019).

\bibitem{Xing2020}
Y. Xing {\it et~al.}, National Science Review {\bf 7},  579  (2020),
  arXiv:1805.10883 [cond-mat].

\bibitem{Shen2020}
D. Shen {\it et~al.}, Communications Materials {\bf 1},  1  (2020).

\bibitem{Song2021}
J. Song {\it et~al.}, Physical Review X {\bf 11},  021065  (2021).

\bibitem{Liu2022}
Q. Liu {\it et~al.}, Technical report,  (unpublished), arXiv:2206.03405
  [cond-mat] type: article.

\bibitem{Butch2011}
N.~P. Butch, P. Syers, K. Kirshenbaum, A.~P. Hope, and J. Paglione, Physical
  Review B {\bf 84},  220504  (2011).

\bibitem{Bay2012}
T.~V. Bay, T. Naka, Y.~K. Huang, and A. de~Visser, Physical Review B {\bf 86},
  064515  (2012).

\bibitem{Tafti2013}
F.~F. Tafti {\it et~al.}, Physical Review B {\bf 87},  184504  (2013).

\bibitem{Majumder2019}
R. Majumder and M.~M. Hossain, Computational Condensed Matter {\bf 21},  e00402
   (2019).

\bibitem{Meinert2016}
M. Meinert, Physical Review Letters {\bf 116},  137001  (2016).

\bibitem{Brydon2016}
P. Brydon, L. Wang, M. Weinert, and D. Agterberg, Physical Review Letters {\bf
  116},  177001  (2016).

\bibitem{Timm2017}
C. Timm, A.~P. Schnyder, D.~F. Agterberg, and P.~M.~R. Brydon, Physical Review
  B {\bf 96},  094526  (2017).

\bibitem{Savary2017}
L. Savary, J. Ruhman, J.~W.~F. Venderbos, L. Fu, and P.~A. Lee, Physical Review
  B {\bf 96},  214514  (2017).

\bibitem{Boettcher2018}
I. Boettcher and I.~F. Herbut, Physical Review Letters {\bf 120},  057002
  (2018).

\bibitem{Wang2018}
Q.-Z. Wang, J. Yu, and C.-X. Liu, Physical Review B {\bf 97},  224507  (2018).

\bibitem{Bahari2022}
M. Bahari, S.-B. Zhang, and B. Trauzettel, Physical Review Research {\bf 4},
  L012017  (2022).

\bibitem{Ishihara2021}
K. Ishihara {\it et~al.}, Physical Review X {\bf 11},  041048  (2021).

\bibitem{Banerjee2015}
A. Banerjee {\it et~al.},  in {\em Bulletin of the {American} {Physical}
  {Society}} (APS March Meeting 2015, ADDRESS, 2015), Vol.~Volume 60, Number 1.

\bibitem{Canfield1991}
P.~C. Canfield {\it et~al.}, Journal of Applied Physics {\bf 70},  5800
  (1991).

\bibitem{Note1}
Supplemental Material and references \cite {Kresse1996, Bloechl1994,
  Perdew1996, Mostofi2008, Franchini2012} therein. Additionally, relevant input
  and output data of the first-principles study can be downloaded from \protect
  \url {https://doi.org/10.5281/zenodo.7352135}.

\bibitem{Liu2016}
Z.~K. Liu {\it et~al.}, Nature Communications {\bf 7},  12924  (2016).

\bibitem{Note2}
The small Fermi surface pocket centred around \(\Gamma \) seen in experiment is
  dominated by strongly dispersing Bi p\(x\) and p\(_y\) orbitals. While it is
  well captured by our DFT calculation, we neglect its contribution to the
  surface low-energy model due to its small density of states at the Fermi
  level.

\bibitem{Note3}
We have checked the convergence of our results at zero temperature by
  comparison with integration meshes containing \(N_i=2000\) points per
  direction.

\bibitem{Scalapino1986}
D.~J. Scalapino, E. Loh, and J.~E. Hirsch, Physical Review B {\bf 34},  8190
  (1986).

\bibitem{Graser2009}
S. Graser, T.~A. Maier, P.~J. Hirschfeld, and D.~J. Scalapino, New Journal of
  Physics {\bf 11},  025016  (2009).

\bibitem{Kemper2010}
A.~F. Kemper {\it et~al.}, New Journal of Physics {\bf 12},  073030  (2010).

\bibitem{Duerrnagel2022}
M. Dürrnagel, J. Beyer, R. Thomale, and T. Schwemmer, The European Physical
  Journal B {\bf 95},  112  (2022).

\bibitem{Kiesel2012}
M.~L. Kiesel, C. Platt, W. Hanke, D.~A. Abanin, and R. Thomale, Physical Review
  B {\bf 86},  020507  (2012).

\bibitem{Nandkishore2012}
R. Nandkishore, L.~S. Levitov, and A.~V. Chubukov, Nature Physics {\bf 8},  158
   (2012).

\bibitem{Scheurer2017}
M.~S. Scheurer, D.~F. Agterberg, and J. Schmalian, npj Quantum Materials {\bf
  2},  1  (2017).

\bibitem{Read2000}
N. Read and D. Green, Physical Review B {\bf 61},  10267  (2000).

\bibitem{Senthil1999}
T. Senthil, J.~B. Marston, and M.~P.~A. Fisher, Physical Review B {\bf 60},
  4245  (1999).

\bibitem{Horovitz2003}
B. Horovitz and A. Golub, Physical Review B {\bf 68},  214503  (2003).

\bibitem{Hecker2018}
M. Hecker and J. Schmalian, npj Quantum Materials {\bf 3},  1  (2018).

\bibitem{Fernandes2019}
R.~M. Fernandes, P.~P. Orth, and J. Schmalian, Annual Review of Condensed
  Matter Physics {\bf 10},  133  (2019).

\bibitem{Note4}
More details are provided in the Supplemental Material and references \cite
  {Polyakov1975, Friedan1980, Orth2014} therein.

\bibitem{Cho2020}
C.-w. Cho {\it et~al.}, Nature Communications {\bf 11},  3056  (2020).

\bibitem{https://doi.org/10.48550/arxiv.2211.13106}
J. Kim {\it et~al.}, Molecular Beam Epitaxy of a Half-Heusler Topological
  Superconductor Candidate YPtBi, 2022.

\bibitem{Kresse1996}
G. Kresse and J. Furthmüller, Physical Review B {\bf 54},  11169  (1996).

\bibitem{Bloechl1994}
P.~E. Blöchl, Physical Review B {\bf 50},  17953  (1994).

\bibitem{Perdew1996}
J.~P. Perdew, K. Burke, and M. Ernzerhof, Physical Review Letters {\bf 77},
  3865  (1996).

\bibitem{Mostofi2008}
A.~A. Mostofi {\it et~al.}, Computer Physics Communications {\bf 178},  685
  (2008).

\bibitem{Franchini2012}
C. Franchini {\it et~al.}, Journal of Physics: Condensed Matter {\bf 24},
  235602  (2012).

\bibitem{Polyakov1975}
A.~M. Polyakov, Physics Letters B {\bf 59},  79  (1975).

\bibitem{Friedan1980}
D. Friedan, Physical Review Letters {\bf 45},  1057  (1980).

\bibitem{Orth2014}
P.~P. Orth, P. Chandra, P. Coleman, and J. Schmalian, Physical Review B {\bf
  89},  094417  (2014).

\bibitem{Note5}
\protect \url {https://doi.org/10.5281/zenodo.7352135}.

\end{thebibliography}

\appendix

\section{Supplementary Material}
\subsection{Density functional theory calculations}
We employed first-principles calculations based on the density functional theory as implemented in the Vienna ab-initio simulation package (VASP)~\cite{Kresse1996}, within the projector-augmented plane-wave (PAW) method~\cite{Bloechl1994}.
The generalized gradient approximation as parametrized by the PBE-GGA functional for the
exchange-correlation potential is used~\cite{Perdew1996} by expanding the Kohn-Sham wave functions into plane waves up to an energy cutoff of \SI{300}{\electronvolt}.
A hexagonal supercell consisting of six (6) unit cells and \SI{22.5}{\angstrom} of vacuum is used to simulate a Bi-terminated crystal. We sampled the Brillouin zone on a dense $16\times16\times1$ regular mesh by including SOC self-consistently.
As stated in the main text, for the calculation of the low-energy Wannier model, the Kohn-Sham wave functions were projected onto a spinful Bi $p_z$ orbital by using the VASP2WANNIER90 interface~\cite{Mostofi2008,Franchini2012}. Relevant input and output data of the first-principles study can be downloaded from~\footnote{\url{https://doi.org/10.5281/zenodo.7352135}}.

\subsection{Vestigial order}
In this section we analyze the emergence of vestigial order above the actual transition temperature.
The superconducting order parameter of our analysis transforms according to the $E$-representation of the point group $C_{{\rm 3v}}$.
The corresponding Ginzburg-Landau expansion of this state is
\begin{equation}\label{eq:GL}
	\begin{split}
		S = &\int d^{2}x\left(r_{0}\boldsymbol{\psi}^{\dagger}\boldsymbol{\psi}+u\left(\boldsymbol{\psi}^{\dagger}\boldsymbol{\psi}\right)^{2}-w\left(\boldsymbol{\psi}^{\dagger}\tau^{y}\boldsymbol{\psi}\right)^{2}\right) \\
		&+S_{{\rm grad}} \mperiod
	\end{split}
\end{equation}
Here, $S_{{\rm grad}}$ contains gradient terms and
\begin{equation}
	\boldsymbol{\psi}=\left(\psi_{1},\psi_{2}\right)^{T}
\end{equation}
summarizes the two components that span the two-dimensional irreducible
representation with basis functions $\Phi_{i}^{E}\left(\boldsymbol{k}\right)\chi_{i\sigma\sigma'}$ composed from the momentum ($\Phi$) and spin ($\chi$) dependence.
We use $\tau^{y}$ to denote the Pauli matrix acting in this two-dimensional space.

The order parameter which determines the pairing amplitude
\begin{equation*}
	\Delta_{\sigma\sigma'}\left(\boldsymbol{x},\boldsymbol{k}\right)=
	\sum_{\boldsymbol{r}}
	e^{i\mathbf{k}\mathbf{r}}
	\left\langle
		c_{\boldsymbol{x}+\frac{\boldsymbol{r}}{2},\sigma}
		c_{\boldsymbol{x}-\frac{\boldsymbol{r}}{2},\sigma'}
	\right\rangle
\end{equation*}
is given as
\begin{equation}
\Delta_{\sigma\sigma'}\left(\boldsymbol{x},\boldsymbol{k}\right)=\sum_{i}\psi_{i}\left(\boldsymbol{x}\right)\Phi_{i}^{E}\left(\boldsymbol{k}\right)\chi_{i\sigma\sigma'} \mcomma
\end{equation}
where $\boldsymbol{x}$ is the center of mass coordinate of the Cooper pair and $\boldsymbol{k}$ denotes it's relative momentum.
The spin dependence of a pure singlet Cooper pair reads $\chi_{\sigma\sigma'}=i\sigma_{\sigma\sigma'}^{y}$.
Despite our findings for \ce{LuPtBi} which we presented in the main text, our present analysis is valid for an arbitrary spin structure, that is, for any mixture of singlet and triplet spin states.

In \autoref{eq:GL} one could also add terms like $\left(\boldsymbol{\psi}^{\dagger}\tau_{x}\boldsymbol{\psi}\right)^{2}$
or $\left(\boldsymbol{\psi}^{\dagger}\tau_{z}\psi\right)^{2}$, but
for $C_{{\rm 3v}}$ they must have the same coefficient and can therefore
be eliminated using a Fierz identity
\begin{equation}
\sum_{\alpha=x,y,z}\left(\boldsymbol{\psi}^{\dagger}\tau_{\alpha}\boldsymbol{\psi}\right)^{2}=\left(\boldsymbol{\psi}^{\dagger}\boldsymbol{\psi}\right)^{2} \mperiod
\end{equation}

One easily finds that the superconductor breaks time-reversal symmetry
with
\[
\varphi\equiv\left\langle \boldsymbol{\psi}^{\dagger}\tau^{y}\boldsymbol{\psi}\right\rangle =-i\left(\psi_{1}^{*}\psi_{2}-\psi_{2}^{*}\psi_{1}\right)\neq0
\]
whenever $w>0$. The lowest energy occurs for
\[
\boldsymbol{\psi}=\frac{\psi_{0}}{\sqrt{2}}\left(1,\pm i\right)
\]
such that $\varphi=\pm\left|\psi_{0}\right|^{2}$ acts as an effective
Ising variable that signals the two degenerate states with broken TRSB.
If $w<0$, the system would instead break a rotation symmetry where
some combination of $\left\langle \boldsymbol{\psi}^{\dagger}\tau^{x}\boldsymbol{\psi}\right\rangle $
and $\left\langle \boldsymbol{\psi}^{\dagger}\tau^{z}\boldsymbol{\psi}\right\rangle $
becomes finite.
For our system with broken inversion symmetry the
Fermi surface is Rashba split.
If the energy scale of this splitting is large compared to the superconducting gap, as it is certainly the case for our system, it always holds that $w>0$.
This statement is in fact more general than the Ginzburg-Landau expansion of \autoref{eq:GL} and valid for an expansion to arbitrary order in the order parameter~\cite{Scheurer2017}.

Considering the regime with $w>0$ but small, we can now analyze the
role of pairing fluctuations. To this end we explicitly spell out the gradient term mentioned above and obtain the effective
action
\begin{equation}
	\begin{split}
		S_{{\rm eff}} = \frac{1}{2T}\int d^{2}x\Big(
			&\left(\partial_{\mu}\theta\right)^{2} +
			\left(\partial_{\mu}\alpha\right)^{2} +
			\sin^{2}\alpha\left(\partial_{\mu}\beta\right)^{2} + \\
			&2\sin^{2}\alpha\partial_{\mu}\beta\partial_{\mu}\theta
			\Big) + S_{{\rm anis}}
	\end{split}
\end{equation}
with temperature $T$ and anisotropy term
\begin{equation}
S_{{\rm anis}}=-\frac{K}{a^{2}T}\int d^{2}x\sin^{2}\left(2\alpha\right)\sin^{2}\beta.
\end{equation}
$K=w\left|\psi_{0}\right|^{2}a^{2}$ is dimensionless and measures the anisotropy that will eventually lead to the discrete symmetry breaking.
Here, we used the parametrization
\begin{equation}
	\boldsymbol{\psi}=\left|\psi_{0}\right|e^{i\theta}\left(\begin{array}{c}
	\cos\alpha\\
	e^{i\beta}\sin\alpha
	\end{array}\right) \mperiod
\end{equation}
The anisotropy is of course a relevant perturbation.
As long as $K\ll1$ the renormalization group flow is nevertheless dominated by the gradient term.
The flow equations for $T$ at $\begin{gathered}K=0\end{gathered}$ up to two loop is given as
\begin{equation}
\frac{dT}{dl}=\frac{1}{\pi}T^{2}+\frac{1}{2\pi^{2}}T^{3} \mperiod
\label{Ricciflow}
\end{equation}
Here, $l$ determines the running cut off scale $\Lambda(l)=\Lambda e^{-l}$. The flow equation~\eqref{Ricciflow} follows from the Wilson-Polyakov renormalization approach of the nonlinear $\sigma$-model~\cite{Polyakov1975}. 
It can most easily be obtained from a two-loop analysis of the Ricci flow using the Friedan scaling approach~\cite{Friedan1980, Orth2014}.
The flow Eq.~\eqref{Ricciflow} goes towards high temperatures, indicating that the essentially isotropic Goldstone-like fluctuations weaken the trend towards order.
At the same time, the anisotropy grows like $K(l)=Ke^{2l}$ and the isotropic scaling stops at $l_{0}=-\frac{1}{2}\log(K)$.
Therefore, the correlation length $\xi(l)=\xi e^{-l}$ takes the value $\xi(l_{0})=\xi\sqrt{K}$.

When the renormalized correlation length $\xi(l_{0})$ reaches the
lattice constant $a$, the anisotropy dominates the flow and the system undergoes an Ising transition to a state with broken time-reversal symmetry.
Thus, we find $\xi_{\rm TRSB}\sim a/\sqrt{K}$.
This transition happens prior to the Berezinskii-Kosterlitz-Thouless (BKT) transition where the superconducting phase acquires a finite stiffness as $\xi_{\rm BKT}\rightarrow \infty$.
To estimate the magnitude of the BKT transition temperature, we use that the stiffness reaches  $\sim \pi/2$, a value of order unity.
The above flow equation for $T$ yields a value of order unity when $l_{\rm BKT}\sim \pi/T$ such that $T_{\rm BKT}\sim \frac{\pi}{2 \log(1/K)}$.
While of the same order, the Ising temperature $T_{\rm TRSB}$ must, as we discussed earlier, be larger than  $T_{\rm BKT}$.
Hence, a state of vestigial Ising order with broken time-reversal symmetry above the superconducting temperature is inevitable, unless both occur at a joint first order transition.

\end{document}